\documentclass[
    ,final            % use final for the camera ready runs
  ]
  {aipproc}

\layoutstyle{6x9}
\usepackage{amsmath, amsthm, amssymb}
\begin{document}

\title{The Complete Spectral Catalog of Bright BATSE Gamma-Ray Bursts}


\author{Yuki Kaneko}{
  address={Universities Space Research Association},
  altaddress={NSSTC, 320 Sparkman Dr. Huntsville, AL 35805}
}

\author{Robert D. Preece}{
  address={University of Alabama in Huntsville},
  altaddress={NSSTC, 320 Sparkman Dr. Huntsville, AL 35805}
}

\author{Michael S. Briggs}{
  address={University of Alabama in Huntsville},
  altaddress={NSSTC, 320 Sparkman Dr. Huntsville, AL 35805}
}

\author{William S. Paciesas}{
  address={University of Alabama in Huntsville},
  altaddress={NSSTC, 320 Sparkman Dr. Huntsville, AL 35805}
}

\author{Charles A. Meegan}{
  address={NASA/Marshall Space Flight Center},
  altaddress={NSSTC, 320 Sparkman Dr. Huntsville, AL 35805}
}

\author{David L. Band}{
  address={NASA/GSFC, Greenbelt, MD 20771 \& JCA/UMBC, Baltimore, MD 21250}
}

\begin{abstract}
We present a systematic spectral analysis of 350~bright GRBs observed with BATSE, 
with high spectral and temporal resolution. 
Our sample was selected from the complete set of 2704 BATSE GRBs, and
included 17 short GRBs. 
To obtain well-constrained spectral parameters, four different photon models 
were fitted and the spectral parameters that best represent each spectrum were 
statistically determined. 
A thorough analysis was performed on 350~time-integrated and 
8459~time-resolved burst spectra.
Using the results, we compared time-integrated and time-resolved spectral 
parameters, and also studied correlations among the parameters and their 
evolution within each burst.
The resulting catalog is the most comprehensive study of spectral properties of 
GRB prompt emission to date, and provides constraints with exceptional 
statistics on particle acceleration and emission mechanisms in GRBs.
\end{abstract}

\maketitle

%%%%%%%%%%%%%%%%%%%%%%%%%%%%%%%%%%%%%%%%%%%%
%% MAINMATTER
%%%%%%%%%%%%%%%%%%%%%%%%%%%%%%%%%%%%%%%%%%%%

\section{Introduction}
Previous spectral studies of Gamma-Ray Burst (GRB) prompt emission have shown 
compelling evidence that a simple emission mechanism cannot entirely account for 
the observed spectra.
However, spectral parameters that provide constraints on GRB emission 
mechanisms often depend on functional forms used to fit the data, as well as 
integration timescales of the spectra.
A consistent, comprehensive spectral study of GRB prompt emission in large 
quantity is therefore crucial to unveiling the nature of GRBs.

A total of 2704 GRBs were observed with the Large Area Detectors (LADs) of
the Burst and Transient Source Experiment (BATSE) on board the 
{\it Compton Gamma-Ray Observatory}, 
in the energy range of $\sim$ 30 -- 1900 keV. 
BATSE provided the largest database currently available of GRBs from a single 
experiment, with good spectral and temporal resolution. 
Many of the BATSE GRBs were bright enough for high-time resolution spectroscopy 
within the bursts as well as time-integrated spectroscopy.
We present here a systematic spectral analysis of 350 bright BATSE GRBs. 
The sample size is more than twice that of the previous BATSE spectral catalog 
\cite{pre00}, which included only time-resolved spectroscopy.
Our analysis included 350 time-integrated spectra and 8459 time-resolved
spectra, and four different photon models were used to fit all the spectra.
The resulting catalog is the largest and most comprehensive study to date 
of GRB prompt emission spectra.

\section{The Analysis}
To assure sufficiently good statistics, we selected the 350 brightest GRBs 
from the entire BATSE database, with either the peak photon flux (256 ms, 
50 -- 300 keV) $>$ 10 photons~s$^{-1}$~cm$^{-2}$ or the total energy fluence 
(30 -- 2000 keV) $> 2 \times 10^{-5}$~ergs~cm$^{-2}$.

One of eight BATSE modules consisted of a LAD and a Spectroscopy Detector (SD). 
In this work, we used only the LAD data, which provided higher sensitivity, 
to make the analysis more consistent throughout. 
In addition, a problem recently identified with the SD response matrices above 
$\sim$ 3 MeV could render the SD data unreliable for spectral analysis at high
energies \cite{kan05}.
The LAD data types used for the analysis are, in order of priority, High-Energy 
Resolution Burst (HERB), Medium Energy Resolution (MER), and Continuous (CONT) 
data \cite[see][for BATSE data types]{pre00}.
All LADs were gain-stabilized with the usable energy range of
$\sim$ 30 -- 1900 keV.
We binned the spectra in time so that each resolved spectrum has signal 
$> 45\sigma$ above background. 
An integrated spectrum is the sum of all the resolved spectra within the burst, 
and most integrated spectra cover the time range of $T_{90}$.

We used four photon models to fit each spectrum. Each photon model consists 
of two, three, four, and five free parameters, respectively in the order shown 
below.

\begin{enumerate}
\item{{\bf Power Law Model (PWRL)} \\[-3ex]
\begin{footnotesize}
\begin{equation*}
f_{\rm PWRL}(E) = A \left(\frac{\textstyle E}{100 {\rm keV}}\right)
   ^{\lambda}
\end{equation*}
\end{footnotesize}
Parameters: $A =$ amplitude (photons~s$^{-1}$~cm$^{-2}$~keV$^{-1}$) and
$\lambda =$ spectral index} \\[-1ex]

\item{{\bf Comptonized Model (COMP)} \\[-2ex]
\begin{footnotesize}
\begin{equation*}
f_{\rm COMP}(E) = A \left(\frac{\textstyle E}{100 {\rm keV}}\right)
   ^{\alpha} 
   \exp{\left(-\frac{\textstyle E(2+\alpha)}{\textstyle E_{\rm peak}}\right)}
\end{equation*}
\end{footnotesize}
Parameters: $A$, $\alpha =$ low-energy spectral index, and 
$E_{\rm peak} = \nu F_{\nu}$ peak energy (keV)} \\[-1ex]

\item{{\bf GRB Model (BAND)} \cite{ban93} \\[-1ex]
\begin{footnotesize}
\begin{equation*}
f_{\rm BAND}(E) = 
\begin{cases} 
   A \left( \frac{\textstyle E}{\textstyle 100 {\rm keV}}\right)^{\scriptstyle \alpha} 
            \exp{\left(
            -\frac{\textstyle E(2+\alpha)}{\textstyle E_{\rm peak}}
            \right)} &
   \text{if~ } E < (\alpha-\beta) \frac{\textstyle E_{\rm peak}}{\textstyle 2+\alpha} 
   \\
   A  \left[
      \frac{\textstyle (\alpha-\beta)E_{\rm peak}}{\textstyle (2+\alpha)100 {\rm keV}}
      \right]^{\alpha-\beta}
            \exp{(\beta-\alpha)} 
            \left(\frac{\textstyle E}{\textstyle 100 {\rm keV}}\right)^{\beta} &
   \text{if~ }  E \geq (\alpha-\beta) \frac{\textstyle E_{\rm peak}}{\textstyle 2+\alpha}
   \end{cases}
\end{equation*}
\end{footnotesize}
Parameters: $A$, $\alpha$, $\beta =$ high-energy spectral index, and
$E_{\rm peak}$} \\[-1ex]

\item{{\bf Smoothly-Broken Power Law Model (SBPL)} \cite{rmfit, ryd99} \\[-1ex]
\begin{footnotesize}
\begin{equation*}
f_{\rm SBPL}(E) = A \left(\frac{\textstyle E}{\textstyle 100 {\rm keV}}\right)^b 
   10^{(a - a_{\rm piv})}
\end{equation*} \\[-5ex]
\begin{align*}
{\rm where~~~~~ }
a &= m \Lambda \ln{\left(
   \frac{\textstyle e^{q} + e^{-q}}{\textstyle 2}\right)}
&a_{\rm piv} &= m \Lambda \ln{\left(
   \frac {\textstyle e^{q_{\rm piv}} + e^{-q_{\rm piv}}}{\textstyle 2}
   \right)} 
&q &= \frac {\textstyle \log {(E/E_{\rm b})}} {\textstyle \Lambda} \nonumber \\
q_{\rm piv} &= \frac {\textstyle \log {(100 {\rm keV}/E_{\rm b})}} 
   {\textstyle \Lambda} 
&m &= \frac{\textstyle \lambda_2 - \lambda_1}{\textstyle 2} 
&b &= \frac{\textstyle \lambda_1 + \lambda_2}{\textstyle 2} 
\end{align*}
\end{footnotesize}
Parameters: $A$, $\lambda_{1}, \lambda_{2} =$ low- \& high-energy 
spectral indices, $E_{\rm b} =$ spectral break energy (keV), and 
$\Lambda =$ break scale (decades of energy)}
\end{enumerate}

\section{The Results}
We fitted the four photon models to each of the 350 integrated 
spectra and 8459 resolved spectra. 
From the overall performance of each model in fitting all spectra, determined by 
the $\chi^2$ of the fits, we found that many spectra were fitted adequately by 
multiple photon models.
We also confirmed that the low-energy index, $\alpha$, of BAND and COMP
models tends to be harder than the low-energy index $\lambda_1$ of SBPL. 

{\bf Model Comparison within Each Spectrum.}
To find the model that best describes each spectrum (referred to as ``BEST"
here), we looked for significant 
improvements in $\chi^2$ when fitted by more complicated models with more 
parameters, starting from the simplest PWRL model.
When the significant improvements were found, we also required the additional 
parameters to be sufficiently constrained, to assure the parameters are 
meaningful and are indeed needed to represent the spectra.
We found that the majority of the spectra required BAND or SBPL. 
COMP was considered BEST in much larger fraction of the resolved spectra than
the integrated ones due to the existence of 
no-high-energy spectra within bursts as well as the lower signal-to-noise ratio.
The distributions of the BEST spectral parameters for both integrated
and resolved spectra are shown in Figure 1.
To account for the difference between the low-energy indices of BAND, COMP, and
SBPL, we use here the ``effective $\alpha$" \cite{pre98,kan06},
the tangential slope in log scale at 25 keV, as the low-energy index where 
BEST is BAND or COMP. 
The effective $\alpha$ and $\lambda_1$ generally agree within 1$\sigma$. 
$E_{\rm peak}$ is the peak energy of the $\nu$F$_{\nu}$ spectrum, regardless 
of the photon model used.
\begin{figure}[b]
  \includegraphics[width=\textwidth]{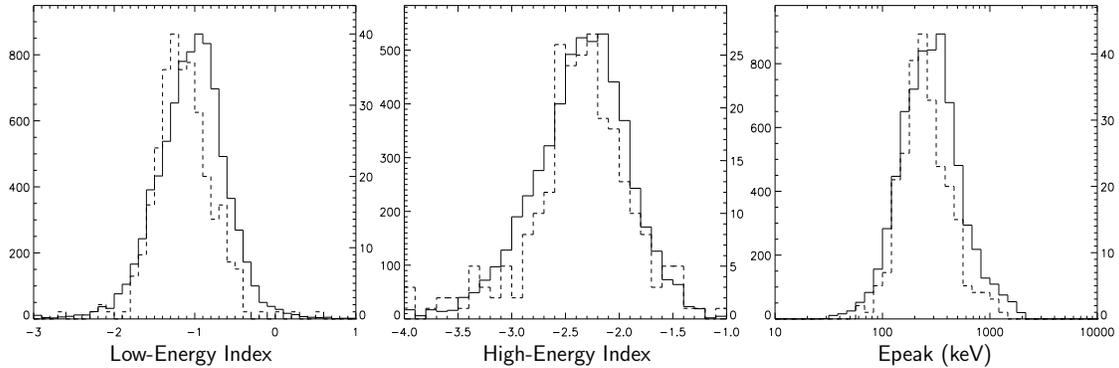}
  \caption{Comparison of BEST parameter distributions of
  integrated (dashed; right axis) and resolved spectra (solid; left axis).
  The last bins include values outside the edge values.}
\end{figure}

{\bf Integrated vs.~Resolved.}
For the comparison between the integrated and resolved parameter distributions
(Figure 1), we calculated the Kolmogorov-Smirnov probabilities and statistics 
($P_{\scriptscriptstyle{\rm KS}}$ and $D_{\rm{\scriptscriptstyle KS}}$) 
for the null hypothesis of two distributions being consistent. 
We found a significant difference ($P_{\scriptscriptstyle{\rm KS}} \sim 10^{-16}$, 
$D_{\rm{\scriptscriptstyle KS}} = 0.23$)
between the low-energy index distributions, and a moderate difference 
($P_{\scriptscriptstyle{\rm KS}} \sim 10^{-2}$, 
$D_{\rm{\scriptscriptstyle KS}} = 0.10$) between their $E_{\rm peak}$ 
distributions. This is due to spectral evolution within bursts, 
commonly observed. 

{\bf Spectral Parameter Evolution \& Correlations.}
An example of the BEST spectral parameter evolution within a burst is
shown in Figure 2.  
In this case, both the low-energy index and $E_{\rm peak}$ evolve from hard to 
soft, and the models with fewer parameters (PWRL, COMP) are BEST at the
tail as the burst spectra become softer. 
\begin{figure}
  \includegraphics[width=\textwidth]{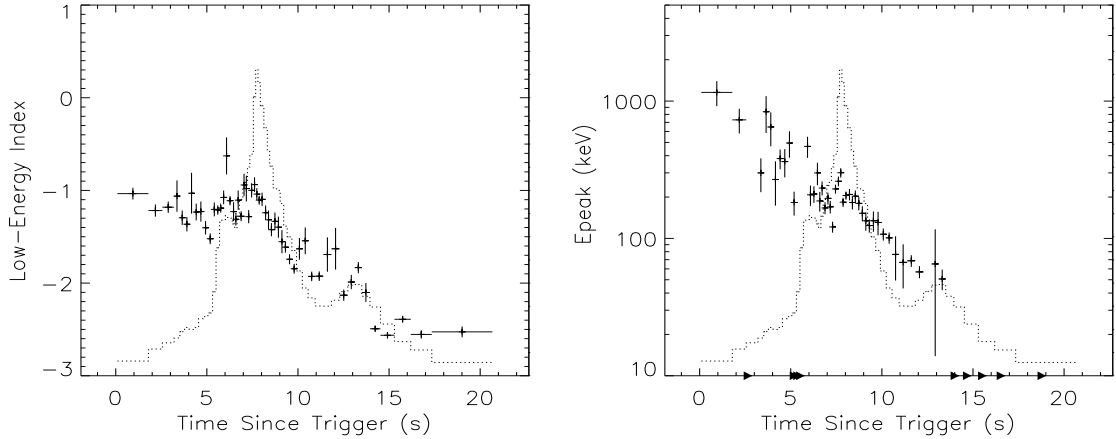}
  \caption{The BEST spectral parameter evolution of GRB 950403.
  Photon flux is overplotted as dotted lines.
  The arrowheads in $E_{\rm peak}$ plot indicate where the values cannot be 
  determined.}
\end{figure}
Within each burst, we also searched for the Spearman rank-order correlations
among the BEST spectral parameters.   
The most significant ($> 3\sigma$) positive correlation was found between 
the $E_{\rm peak}$ and the low-energy index in the largest fraction of GRBs 
(26\%).

{\bf Short GRBs.}
Our sample included 17 short GRBs ($T_{90} <$ 2 s), three of which were
bright enough for time-resolved spectral analysis. Within these
three short GRBs, spectral evolution (either hard-to-soft or photon-flux
tracking behavior) was clearly observed. However, we did
not find any significant differences in the spectral parameters
between the 17 short GRBs and long GRBs in our sample.
A possible reason for this is the fact that only bright GRBs are
considered here, which tend to be harder \cite{mal95}. 

\section{Conclusions}
The GRB spectral database obtained in this work is derived from
the most sensitive and largest database currently available. 
Therefore, these results set a standard for spectral properties of GRB
prompt emission with exceptional statistics. 
Our results can provide reliable constraints for existing and future theoretical 
models of GRB emission and particle acceleration mechanisms.

\bibliographystyle{aipproc}   % if natbib is available

\begin{thebibliography}{8}
\expandafter\ifx\csname natexlab\endcsname\relax\def\natexlab#1{#1}\fi
\providecommand{\enquote}[1]{``#1''}
\expandafter\ifx\csname url\endcsname\relax
  \def\url#1{\texttt{#1}}\fi
\expandafter\ifx\csname urlprefix\endcsname\relax\def\urlprefix{URL }\fi
\providecommand{\eprint}[2][]{\url{#2}}

\bibitem[{Preece, R.D., et al}(2000)]{pre00}
{Preece, R.D., et al}, \emph{ApJS} \textbf{126}, 19 (2000).

\bibitem[{Kaneko, Y.}(2005)]{kan05}
{Kaneko, Y.}, Ph.D. thesis, University of Alabama in Huntsville (2005).

\bibitem[{Band, D.L., et al}(1993)]{ban93}
{Band, D.L., et al}, \emph{ApJ} \textbf{413}, 281 (1993).

\bibitem[{Mallozzi, R.S., Preece, R.D., \& Briggs, M.S.}(1994)]{rmfit}
{Mallozzi, R.S., Preece, R.D., \& Briggs, M.S.}, {{\MakeUppercase WINGSPAN,
  RMFIT}} (1994).

\bibitem[{Ryde, F.}(1999)]{ryd99}
{Ryde, F.}, \emph{ApL\&C} \textbf{39}, 281 (1999).

\bibitem[{Preece, R.D., et al.}(1998)]{pre98}
{Preece, R.D., et al.}, \emph{ApJ} \textbf{506}, L23 (1998).

\bibitem[{Kaneko, Y. et al.}(2006)]{kan06}
{Kaneko, Y. et al.}, \emph{ApJS}  (2006), submitted.

\bibitem[{Mallozzi , R.S., et al.}(1995)]{mal95}
{Mallozzi , R.S., et al.}, \emph{ApJ} \textbf{454}, 597 (1995).

\end{thebibliography}

\end{document}